\documentclass[12pt]{article}
\usepackage{amsmath}
\usepackage[cp1251]{inputenc}  
\usepackage[english,russian]{babel} 
\usepackage{amsfonts,amssymb}
\usepackage[dvips]{graphicx}
\usepackage{psfrag}

\newcommand{\bes}{\begin{equation} }
\newcommand{\ees}{\end{equation} }

\newcommand{\be}{\medskip\begin{equation}}
\newcommand{\ee}{\end{equation}  \vskip4pt}

\begin{document}
\centerline{\Large\bf The properties of Vlasov--Maxwell--Einstein  equations}

\centerline{\Large\bf and its applications to cosmological models}

\vskip 10pt
\centerline{\large\bf Victor Vedenyapin, Nikolay Fimin and Valery  Chechetkin}

\vskip 10pt
\centerline{\small Keldysh Institute of Applied Mathematics of RAS, 125047, Miusskaya sq., 4, Moscow, Russia}

\vskip 10pt

{{\bf Abstract.}\:
The method of obtaining  of Vlasov--type equations for systems of interacting massive charged particles
from the general relativistic  Einstein--Hilbert action is considered.
An effective approach to synchronizing the proper times of various particles of a many--particle system is proposed.
Based on the resulting expressions for the relativistic actions,
an  analysis of composite structure of cosmological term in Einstein's equations is performed.
%
%
%
\section{Introduction}
\label{intro}
Based on the classic   Maxwell--Einstein--Hilbert   action \cite{Pauli}--\cite{Landau2},
we can obtain the Vlasov--Einstein  and Vlasov--Maxwell--Einstein equations and its post--Newtonian approximations with uniform way.
In this case, first we variate the trajectories of the particles,
getting the equations of motion, and then, with the help of the distribution functions formalism, we introduce the Liouville equation.
After that we variate the fields,
for which we preliminarily rewrite actions with using of distribution functions.

In the case of the Vlasov--Einstein--Maxwell equations, new difficulties arise:
this  case requires synchronization of the times of different particles and
comparison of different forms of Lagrangians for geodesics. An integral of the interval appears, which
was usually assumed to be unity  \cite{Pauli}--\cite{5},
without which
it is impossible to synchronize the times of different particles, and therefore to write the Vlasov--Einstein
equation for multiparticle system.
To get the equations of self--consistent fields it requires the conversion of classical actions  from Lagrangian coordinates to Eulerian using
distribution functions.

We  can illustrate this approach  for simple situations of  weak relativistic systems. Then
we get the possibility to analyze the cosmological term in the equations of the General Theory of Relativity,
getting expressions that lead to the same mathematical
conclusions as empirically introduced in the field equations the cosmological term.
Based on this fact, we  conclude  that: 1)dark matter and dark energy  may connect  with the cosmic plasma;
2)antigravity, as an attribute
of dark energy, may connect  with
electrostatic repulsion (so, we can eliminate from consideration any other long--range interaction
for interpretation of antigravity).

\section{Derivation of the Liouville equation in an extended $9$--dimensional phase space}
\label{sec:1}

Relativistic action for moving
 charged (with charge $e$) particles of mass $m$ in the presence of a gravitational and electromagnetic field
 can be written as follows:
 $$
 S_{1}= -mc \int^{\lambda_{max}}_0
 \sqrt{g_{\mu \nu}({\bf X})\frac{dX^\mu(\lambda)}{d\lambda}\frac{dX^\nu(\lambda)}{d\lambda}}d\lambda -
 \frac{e}{c}\int A_\mu \frac{dX^\mu}{d\lambda}d\lambda,
 $$
\noindent
where: $g_{\mu \nu}({\bf X})$ is a metric tensor of $4$--dimensional space--time
(${\bf X} = \{ X^\mu \}_{\mu = \overline{0,3}}$),
$A_\mu ({\bf X}) \equiv \{ \varphi ({\bf X}); {\bf A}({\bf X}) \}$ is a $4$--potential of electromagnetic field;
variable  $\lambda \in {R}^+$   is proportional to individual time of particle (i. e. affine parameter):
$ds = \sqrt{I}d\lambda$, $I \equiv g_{\mu \nu}
(dX^\mu/d\lambda ) (dX^\nu/d\lambda )$ (the physical sense of coefficient $\sqrt{I}$ we'll consider below).

Let's introduce also the action with the modified first term:
$$
S_{2}= -\frac{mc}{2\sqrt{I}}\int^{\lambda_{max}}_0
{g_{\mu \nu}({\bf X})\frac{dX^\mu({\bf q},\lambda)}{d\lambda}\frac{dX^\nu({\bf q},\lambda)}{d\lambda}}
d\lambda
-
\frac{e}{c}\int A_\mu \frac{dX^\mu}{d\lambda}d\lambda.
$$
In literature similar operation (transition to a new form of action)
is made for case an electromagnetic field (the second term in actions $S_{1,2}$),
and is justified by the fact that the equations of motion of a particle
in a gravitational field
will be the same in both
cases (i. e.  when using
actions $S_1$ and $S_2$ with replacing the parameter $\lambda$ with the ``natural''
parameter $s$ or the proper time $\tau = s / c$).

We consider the question of substantiating equivalence of actions on the basis of
coincidences of the Euler--Lagrange equations.
Consider two types of actions with kernels (Lagrangians) of the following general form:
$$
S_I = k \int L \big( {\bf X}, \frac{d {\bf X}}{d\lambda}  \big)d\lambda + \int L_1 \big( {\bf X}, \frac{d {\bf X}}{d\lambda}  \big)d\lambda,
$$
$$
S_{II} =  \int h(L) d\lambda + \int L_1 \big( {\bf X}, \frac{d {\bf X}}{d\lambda}  \big)d\lambda,
$$
\noindent
where $h (L)$ is some (smooth) arbitrary function of its argument.
Let us compare the Euler--Lagrange equations obtained from the actions of $S_I$
and $S_ {II}$.

{\it Lemma~1} (on the equivalence of the actions of $S_I$ and $S_ {II}$).
The sufficient conditions for the equivalence of the actions of $S_I$ and $S_ {II}$
(in the sense of the coincidence of the Euler--Lagrange equations) have the following form:

1)the Lagrangian $L \big({\bf X}, {\bf X}_{\lambda} \big) $ should be the integral of the motion for the action $S_I$;

2)the coefficient $k$ in the definition of $S_I$ must coincide with the derivative of the function $h(L)$ from
definitions of the action of $S_{II}$: $k = dh(L)/dL$.
If the Lagrangian is not equal to zero, then the coefficient $k$ is uniquely determined.

{\it Proof}~ is obtained by directly varying the action of $S_{II}$ generating the equations
    Euler--Lagrange:
  $$
 \frac{d^2 h}{dL^2}  \frac{dL}{d\lambda} \frac{\partial L}{\partial {\bf X}_\lambda} +
 \frac{d h}{dL}   \frac{d}{d\lambda}  \frac{\partial L}{\partial {\bf X}_{\lambda}} +
 \frac{d}{d\lambda} \frac{\partial L_1}{\partial {\bf X}_{\lambda}} =
 \frac{d h}{dL} \frac{\partial L}{\partial {\bf X}} + \frac{\partial L_1}{\partial {\bf X}},
  $$
  \noindent
  and comparing the resulting equations with the corresponding equations of motion for the action of $S_I$:
  $$
k \frac{d}{d\lambda}  \frac{\partial L}{\partial {\bf X}_{\lambda}}     +  \frac{d}{d\lambda}  \frac{\partial L_1}{\partial {\bf X}_{\lambda}} =
k \frac{\partial L}{\partial {\bf X}}  +  \frac{\partial L_1}{\partial {\bf X}}.
$$
A consequence of this Lemma is the fact that the previously introduced actions $S_1$ and $S_{2}$
  are equivalent in the sense of the Lemma, that is, they have
  identical equations of motion. Indeed, for these actions we have
 $$
 h(L)= -mc\: \sqrt{L},~\:~~ L= g_{\mu \nu}\frac{dX^\mu}{d\lambda} \frac{dX^\nu}{d\lambda},~~~~
 L_1=-\frac{e}{c}A_\mu \frac{dX^\mu}{d\lambda}.
 $$
 Condition 1) of the Lemma are satisfied by the Euler homogeneous function theorem:
 the Hamilton function (integral of motion!) for the action $S_2$
(obtained by application
the Legendre transform is proportional to the Lagrangian $L = g_{\mu \nu} {X^\mu}_\lambda {X^\nu}_\lambda$), and the Lagrangian
$L_1$ are the 1st degree
  by the ``velocity'' variable ${X^\mu}_\lambda $; condition 2) is satisfied since the coefficient $k$ in $S_I$ is equal to
   derivative of
   function $h(L)$  (from definition of action $S_{II}$): $k = dh / dL = -mc / (2 \sqrt{L})$.
   The value  $I$ is numerically equal to
   value of the Lagrangian $L$ (and is proportional to the corresponding Hamiltonian).

We write  the Euler--Lagrange equations for the actions $S_1$ or $S_2$. In accordance with the Lemma,
they are identical
when varying
$S_1$
(the interval value is assumed to be not equal to unity, but $\sqrt{I}$):
\begin{equation}
\frac{mc}{\sqrt{I}}\frac{d}{d\lambda}\bigg(
g_{\mu \nu}  \frac{dX^\nu}{d\lambda}
\bigg) + \frac{e}{c} \frac{dA_\mu}{d\lambda} =
\frac{mc}{2\sqrt{I}}  \frac{\partial g_{\nu \zeta}}{\partial X^\mu}  \frac{dX^\nu}{d\lambda}   \frac{dX^\zeta}{d\lambda}
+ \frac{e}{c}  \frac{\partial A_\nu}{\partial X^\mu}  \frac{dX^\nu}{d\lambda}.
\label{6dan}
\end{equation}
This shows that in the absence of electromagnetic interaction between particles,
the quantity $mc / \sqrt{I}$ is reduced, and
the equations of motion are the same using both the $\lambda$ parameter and the
 interval parameter $s$.
However, taking into account the electromagnetic interaction leads to different equations when
using various parameters. Although how
can be seen from the equation (\ref{6dan}), it is possible in principle to transfer to the affine parameter $s$,
expressing $d \lambda$ in terms of $ds$ and $I$:
$ds = \sqrt{I} d \lambda$.

In multiparticle systems, this is not possible. Consider an action similar to $S_1$, but for a system of many particles with
with different masses $ m_{a} $ and charges $ e_{a} $ ($a = \overline{1, N}$):
$$
S_{1, \Sigma} = -\sum_{a}   m_{a}c \int \sqrt{g_{\mu \nu} \frac{dX^\mu_{a}}{d\lambda}
\frac{dX^\nu_{a}}{d\lambda}}d\lambda
-\sum_{a} \frac{e_{a}}{c}\int A_\mu \frac{dX^\mu_{a}}{d\lambda} d\lambda.
$$
Again, we transfer to the Lagrangian quadratic in velocity, and we obtain the equivalent action:
$$
S_{2, \Sigma} = - \sum_a  \frac{m_a c}{2 \sqrt{I_a}} \int g_{\mu \nu}
({\bf X}_a) \frac{dX^\mu_a}{d\lambda}  \frac{dX^\nu_a}{d\lambda}d\lambda
- \sum_a \frac{e_a}{c} \int A_\mu \frac{dX_a^\mu}{d\lambda}d\lambda.
$$
We note here the appearance of the index $a$ (numerating the particles) in the integral $I_a$: the values of these integrals,
denoting the size of the interval of different particles are not necessarily the same.
By this we synchronized the proper time of different particles $ds_a =\sqrt{I_a} d\lambda$
in the following sense: 1)we found that the impossibility of synchronizing the $ ds_a $ intervals themselves is related
with various values of the integrals $I_a$; 2)we demonstrated how different proper times are related: the  parameter $\lambda$
for all particles is the same. Note that the integrals $I_a$ depend on the parameterization, but their ratio is not
depends on ($I_{a_1} / I_{a_2} \neq \phi (\lambda),~ a_{1,2} \in \{1, ..., N \}$).

To describe the dynamics of a many--particle system associated with the actions of $S_{1, \Sigma}$ or $S_{2, \Sigma}$,
canonical (``long'') momenta can be introduced in a standard way:
$$
(Q_{a})_\mu = \frac{\partial L}{\partial V_a^\mu} = - \frac{m_a c}{\sqrt{I_a}} g_{\mu \nu} ({\bf X}_a) V_a^\nu
- \frac{e_a}{c}A_\mu ({\bf X}_a),~~~V_a^\nu \equiv \frac{\partial X_a^\nu}{\partial \lambda}.
$$
Obviously, we can get an explicit expression of the velocities through canonical momenta:
$$
V_a^\nu = - \frac{\sqrt{I_a}}{m_a c}    g^{\mu \nu} ({\bf X}_a) \big( (Q_{a})_\mu +  \frac{e_a}{c}A_\mu  \big).
$$
Accordingly, the second equation of the Hamiltonian pair of equations associated with canonically conjugate variables
$({\bf X}_a, {\bf Q}_a)$:
$$
\frac{d (Q_{a})_\mu}{d\lambda} = \sum_a \frac{\sqrt{I_a}}{m_a c}  \big( (Q_{a})_\zeta +
 \frac{e_a}{c}A_\zeta  ({\bf X}_a) \big) \frac{\partial g^{\zeta \nu}}{\partial X^\mu_a}
 \big( (Q_{a})_\nu +  \frac{e_a}{c}A_\nu ({\bf X}_a)  \big)  +
$$
$$
+  \frac{e_a \sqrt{I_a}}{m_a c^2}\big( (Q_{a})_\zeta + \frac{e_a}{c}A_\zeta
 ({\bf X}_a)\big)g^{\zeta \xi}\frac{\partial A_\xi ({\bf X}_a)}{\partial X_a^\mu}.
$$
Moreover, the Hamilton function corresponding to these equations has the form:
$$
H = \sum_a  \frac{\sqrt{I_a}}{m_a c}
\big( (Q_{a})_\zeta + \frac{e_a}{c}A_\zeta  ({\bf X}_a) \big)  g^{\zeta \nu}
\big( (Q_{a})_\nu + \frac{e_a}{c}A_\nu  ({\bf X}_a) \big).
$$
Here the integrals $\sqrt{I_a}$ synchronize the times,
leading to differentiation with respect to the same parameter $\lambda $:
the relation $ds_a = \sqrt{I_a} d\lambda$ demonstrates that equations
are obtained where one can go to proper (generally speaking, different) times.
We introduce (partial, for the type of $a$ particles)
the distribution function $f_a \big({\bf X}, {\bf Q}, \lambda \big)$ over the extended 9--dimensional
phase space
(the indices $a$ have moved from coordinates and momenta to the distribution function $f_a$).
The Liouville equation for $f_a$ takes the following form:
\begin{equation}
\frac{\partial f_a \big({\bf X}, {\bf Q}, \lambda  \big)}{\partial \lambda}  - \frac{\sqrt{I_a}}{m_a c}  g^{\mu \nu}({\bf X}_a)
\big(  (Q_{a})_\mu   + \frac{e}{c}A_\mu  \big) \frac{\partial f_a}{\partial X^\nu} \,+
\label{12dan}
\end{equation}
$$
+\: \bigg( \frac{\sqrt{I_a}}{m_a c}  \big( (Q_{a})_\zeta + \frac{e_a}{c}A_\zeta  ({\bf X}_a) \big)
\frac{\partial g^{\zeta \nu}}{\partial X^\mu_a}
\big( (Q_{a})_\nu +  \frac{e_a}{c}A_\nu ({\bf X}_a)  \big)  +
$$
$$
+\: \frac{e_a \sqrt{I_a}}{m_a c^2} \big(  (Q_{a})_\zeta + \frac{e_a}{c}A_\zeta  ({\bf X}_a) \big)
g^{\zeta \xi} \frac{\partial A_\xi}{\partial X^\mu_a}\bigg)
\frac{\partial f_a}{\partial Q_\mu}=0.
$$
The equations depend on the index $a$ through the masses $m_a$, the charges $e_a$ and  integrals $I_a$. Let us write
  $\lambda$--stationary form of this equation,
when $f_a$ does not depend on the parameter $\lambda$ (in similar form the Vlasov--Einstein equation is usually written in
literature, although for simplified case of absence  of
electromagnetic interaction in multiparticle system):
$$
-  g^{\mu \nu}({\bf X}_a) \big(  (Q_{a})_\mu   + \frac{e}{c}A_\mu  \big) \frac{\partial f_a ({\bf X},{\bf Q})}{\partial X^\nu} \,+\,
\bigg( \frac{\partial g^{\zeta \nu}}{\partial X^\mu_a} \big( (Q_{a})_\zeta + \frac{e_a}{c}A_\zeta  \big)
\big( (Q_{a})_\nu +  \frac{e_a}{c}A_\nu  \big) \, +
$$
$$
\,+\,  \frac{e_a}{c}F_{\mu \nu} ({\bf X})  g^{\zeta \nu} \big(  (Q_{a})_\zeta + \frac{e_a}{c}A_\zeta  \big) \bigg)
\frac{\partial f_a}{\partial Q_\mu}=0.
$$
We can compare the kinetic equations written above with the Liouville equations, where
noncanonical (``short'') momenta
    with zero electromagnetic fields in action $S_{1, \Sigma}$: $(P_a)_\mu = -m_a c I^{-1/2}_a g_{\mu \nu}
({\bf X}_a) V_a^\nu$.
     The resulting equations are non--Hamiltonian, but divergent--free:
  \begin{equation}
   \frac{dX_a^\nu}{d\lambda}  = - \frac{\sqrt{I_a}}{m_a c}g^{\mu \nu} ({\bf X}) (P_a)_\mu,
   \label{26dan}
   \end{equation}
   $$
    \frac{d (P_a)_\mu}{d\lambda}  = - \frac{\sqrt{I_a}}{m_a c} \frac{\partial g^{\nu \zeta}}{\partial X^\mu} (P_a)_\nu  (P_a)_\zeta  +  \frac{e_a}{c}
    \frac{\sqrt{I_a}}{m_a c} F_{\mu \nu} ({\bf X}_a)  g^{\zeta \nu} ({\bf X}_a)  (P_a)_\zeta.
   $$
   We note that is the  similar situation with time synchronization of particles:
proper times all differ, as the formula  $ds_a = \sqrt{I_a} d \lambda$ demonstrates.

Let us write  the Liouville equation,
introducing the partial distribution functions of $f_a ({\bf X}, {\bf P}, {\lambda})$  of the
particles with masses $m_a$ and charges $e_a$
over a 9--dimensional phase space $({\bf X}, {\bf P}, \lambda)$:
$$
\frac{\partial f_a  ({\bf X},{\bf P}, \lambda)}{\partial \lambda} - \frac{\sqrt{I_a}}{m_a c} g^{\mu \nu}
({\bf X}) (P_a)_\mu \frac{\partial f_a}{\partial X^\nu}\,+ $$
$$+\,
\bigg(
-\frac{\sqrt{I_a}}{m_a c} \frac{\partial   g^{\nu \zeta}}{\partial X^\mu}(P_a)_\nu (P_a)_\zeta
+  \frac{e_a}{c} \frac{\sqrt{I_a}}{m_a c} F_{\mu \nu} ({\bf X})  g^{\zeta \nu}(P_a)_\zeta
\bigg)\frac{\partial f_a}{\partial P_\mu}=0.
$$
This equation can be rewritten in a form that excludes the parameter  $\lambda$,
if we replace this parameter with fixed interval of
$a_0$--th particle ($a_0 \in \{1, ..., N \}$)
according to the formula $d \lambda = ds_{a_0} / \sqrt {I_{a_0}}$:
$$
\frac{\partial f_a  ({\bf X},{\bf P}, s)}{\partial s_{a_0}} - \frac{1}{m_a c} \frac{\sqrt{I_{a}}}{\sqrt{I_{a_0}}}
g^{\mu \nu} ({\bf X}) (P_a)_\mu \frac{\partial f_a}{\partial X^\nu}\,+ $$
$$+\,
\bigg(
-\frac{1}{2m_a c}\frac{\sqrt{I_{a}}}{\sqrt{I_{a_0}}} \frac{\partial   g^{\nu \zeta}}{\partial X^\mu}(P_a)_\nu (P_a)_\zeta
+  \frac{e_a}{c} \frac{1}{m_a c}\frac{\sqrt{I_{a}}}{\sqrt{I_{a_0}}} F_{\mu \nu} ({\bf X})  g^{\zeta \nu}(P_a)_\zeta
\bigg)\frac{\partial f_a}{\partial P_\mu}=0.
$$
Moreover, as we noted above, the ratio $I_a / I_{a_0}$ does not depend on $\lambda$
  (and it is a function of the variable ${\bf X}$  only).

We consider the $\lambda$--stationary
form of the Liouville equation when $f_a = f_a ({\bf X}, {\bf P})$,  $f_a$  does not depend
  from the parametric variable $\lambda $ (while
the factors $\sqrt{I_a} / (m_a c)$ on the left side of the equation are reduced):
$$
-g^{\mu \nu} ({\bf X})P_\mu \frac{\partial f_a ({\bf X},{\bf P})}{\partial X^\nu} + \bigg(-\frac{1}{2}
\frac{\partial   g^{\nu \zeta}}{\partial X^\mu}P_\nu P_\zeta
+  \frac{e_a}{c} F_{\mu \nu} ({\bf X})  g^{\zeta \nu}P_\nu
\bigg)\frac{\partial f_a}{\partial P_\mu}=0
$$
\noindent
(since $X^0 = ct$, the last equation is not  $t$--stationary  in the general case). This type of equations
is usually considered as the Vlasov--Einstein equations \cite{sad1}--\cite{sad4}.
But this  type of equations  does not conserve particle number, as following  Lemmas show.

We consider more  general question of the conservation of the number of particles for the Liouville equation in time.
Usually, in works devoted to
Vlasov--Einstein equation
 authors prove the conservation of the particles   number  $\int f |g| d^4X d^4V$, but this is not enough.
We illustrate the difficulties 0f
proofs with using the following two Lemmas.

{\it Lemma 2} (on the conditions for the balance of phase points in a dynamical system). Let's consider
partial linear equation over the $(N + 1)$--dimensional phase space (${\bf y} = \{y_0, y_1, ..., y_N \}$):
 \begin{equation}
 \frac{\partial f({\bf y})}{\partial y_0}
 +F^i({\bf y})\frac{\partial f}{\partial y^i}+\psi ({\bf y})f({\bf y})=0,~~~i=\overline{1,N}.
 \label{Lemma2}
\end{equation}
This equation preserves the number of points in the phase space if and only if
  $\psi ({\bf y}) = {\rm div}_N \: {\bf F} \equiv \partial F^i / \partial y^i$.

{\it Proof}. We rewrite the equation as follows:
$$
  \frac{\partial f({\bf y})}{\partial y_0}
 +\frac{\partial  (F^i({\bf y})f)}{\partial y^i}+ \big(\psi ({\bf y}) - {\rm div}_N\:{\bf F} ({\bf y}) \big)f({\bf y})=0.
 $$
After integrating this equation over the variable ${\bf y}$,
we obtain the conservation equation for the number of points in the phase space:
$$
\frac{d N}{d y_0}=\int \big({\rm div}_N\:{\bf F} ({\bf y}) -\psi ({\bf y}) \big)f({\bf y})d{\bf y},~~~N({y}_0)
 = \int f({\bf y})dy_1...dy_N.
$$
Thus, the equation $dN /dy_0 = 0$ is equivalent to the condition $\psi = {\rm div}_N\:{\bf F}$ and the Liouville equation
takes the form
$\partial f/\partial y_0 + \partial ({F^i}f)/\partial {y^i}=0$.

{\it Lemma 3} (on the stationary Liouville equation). We will consider an autonomous ODE system
(or a dynamic system):
$$\frac{dy^\mu}{d \lambda} = F^\mu ({\bf y}), ~~~ \mu = 1,...,N.
$$
We associate with this dynamic system the kinetic equation for the density function $f(\lambda, {\bf y})$
over $(N + 2)$--dimensional phase space
$Y^{N + 2} \ni \widetilde {\bf y} $, $ \widetilde {\bf y} = \{\lambda, {\bf y } \}$:
\begin{equation}
 \frac{\partial f}{\partial \lambda} + \frac{\partial }{\partial y^\mu}\big(  {F}^\mu f   \big)=0.
  \label{posadov1}
 \end{equation}
This equation preserves the number of particles in the variable $y_0$, i.e. $d / dy_0 (\int f d^ny) = 0$,
if and only if $({\bf F}, \nabla_{\bf y} F_0) = 0$.

{\it Proof}.
We rewrite the last equation as follows:
$$
 \frac{\partial f}{\partial y_0} + \frac{F^i}{F^0}\frac{\partial f}{\partial y^i}+ \frac{1}{F^0}
  \bigg(  \frac{\partial F^i}{\partial y^i} +  \frac{\partial F^0}{\partial y^0}\bigg)f=0.
$$
Applying Lemma 2 to this equation, we see that the identity $ (\nabla_{\bf y} F^0, {\bf F}) = 0$ is true, i.e., $F^0$
is an integral of motion in the process of evolution of the original dynamic system.

So, we have obtained the condition when the stationary equation preserves the number of particles in time.
This condition is not satisfied for all forms of the previously proposed Vlasov--Einstein equations,
based on the stationary forms of equation (2).
Whole construction
also has the disadvantage that it all depends on the choice of the lambda parameter
or proper time of some particle. Additional troubles for $4$--dimensional
pulses arise when deriving equations for fields where it is necessary to proceed to integration over the mass surface.
Conclusion: in the $4$-dimensional form, the Vlasov--Einstein equation
does not exist, although the name is available.
All this indicates the need to use $3$--dimensional speeds, which will be done in the next paragraph.
And now, all the same, we show the usefulness of equation (3).

{\it Example 1.}\,Consider the special case of the equation (\ref{6dan}), when the metric $g_{\mu \nu}$ and
  components of the vector potential $A_\mu$
independent of the time coordinate. Then the right-hand side of the equality (\ref{6dan}) with
the index $\mu = 0$ is canceled, and possibly
analytically integrate the left side (index $a$ omits):
$$
\frac{mc}{\sqrt{I}}\big(  g_{0 \nu} \frac{dX^\nu}{d\lambda} \big) + \frac{e}{c}A_0 =-Q_0.
$$
The meaning of the resulting integral can be clarified by taking the post--Galilean metric
$$g_{\mu \nu} = {\rm{diag}}\big( 1+2\Phi/c^2,-1,-1,-1   \big)$$
\noindent({\it Landau metric}),
where $\Phi (X^j) $ is the Newtonian gravitational potential.
Then the last ratio is converted to form
\begin{equation}
\frac{mc}{\sqrt{I}}\big(  1+\frac{2\Phi}{c^2}\big) \frac{dX^0}{d\lambda}   + \frac{e}{c}A_0 =-Q_0,
\label{19dan}
\end{equation}
\noindent
and the remaining Euler--Lagrange equations of the system (\ref{6dan}) take the form:
\begin{equation}
\frac{mc}{\sqrt{I}}\frac{d}{d\lambda} \frac{dX^j}{d\lambda} +
\frac{e}{c}
 \frac{dA_j}{d\lambda} = \frac{mc}{2c^2 \sqrt{I}} \frac{\partial \Phi}{\partial X^j} \bigg(
\frac{dX^0}{d\lambda}
\bigg)^2 + \frac{e}{c} \frac{\partial A_\nu}{\partial X^j} \frac{dX^\nu}{d\lambda},~~~j=1,2,3.
\label{EuLag}
\end{equation}

Replacing
the parameter $\lambda $ from the equation (\ref{EuLag}) for the time $t$,
we get the equations of motion
of a charged particle in an electrostatic field and in the gravitational potential $\Phi $:
$$
\frac{d}{dt} \bigg(M \frac{d X^j}{dt} \bigg) = -M \frac{\partial \Phi}{\partial X^j} + \frac{e}{c}F_{\mu j} \frac{d X^\mu}{dt},
$$
where $M = -(Q_0/c - eA_0/c^2)/(1+2\Phi/c^2)$
is an effective mass of a particle in a superposition of fields.
Thus, the effective mass $M$
depends on the gravitational and electromagnetic fields, and in this expression the quantity $Q_0$ can be considered as
as the zero component of the momentum in the absence of external fields. We give an explicit expression for $Q_0$ and for $M$:
$$
Q_0 = -\frac{mc  (1 + 2\Phi/c^2)}{\sqrt{1-{\bf v}^2/c^2 + 2\Phi/c^2}} - \frac{e}{c}A_0,~~~\:
 M= \frac{m}{\sqrt{1-{\bf v}^2/c^2 + 2\Phi/c^2}}.
$$
In addition to the Landau metric, when  we transferred to the
post--Newtonian approximation, we can also consider the {\it Fock metric}:
$$
g_{\mu \nu} = {\rm{diag}}\big( 1+2\Phi/c^2,-(1-2\Phi/c^2),-(1-2\Phi/c^2),-(1-2\Phi/c^2)   \big).
$$
The equation of motion in this case takes the following form:
$$
\frac{d}{dt} \bigg(M \frac{d X^j}{dt} \bigg) = -M  \frac{1+{\bf v}^2/c^2}{1-2\Phi/c^2} \frac{\partial \Phi}{\partial X^j} +
\frac{e}{c}F_{\mu j} \frac{d X^\mu}{dt},
$$
the explicit expressions for $Q_0$ and $M$ (in accordance with \cite{pershin1}--\cite{pershin2}) are :
$$
Q_0 = -\frac{mc  (1 + 2\Phi/c^2)}{\sqrt{1-{\bf v}^2/c^2 + 2\Phi/c^2 + 2\Phi {\bf v}^2/c^4} } - \frac{e}{c}A_0,
$$
$$
M = - \frac{(Q_0/c -
eA_0/c^2)(1-2\Phi/c^2)}{1+2\Phi/c^2} = \frac{m(1-2\Phi/c^2)}{\sqrt{1-{\bf v}^2/c^2 + 2\Phi/c^2 + 2\Phi{\bf v}^2/c^4}}.
$$
We have demonstrated the usefulness of the derived equations (3) before moving on to equations (7).

{\it Example 2.}\,Let us consider the case when gravitational and electromagnetic fields depend only on the time variable $t$
(which means that the universe is completely homogeneous). In this case, equations (1) can be integrated
Hamiltonian mechanics methods. It is interesting to analyze some particular aspects of the situation.
We have here three integrals of motion
$$
\frac{mc}{\sqrt{I}}\big( g_{k\mu} \frac{dX^\mu}{d\lambda}  \big)+\frac{e}{c}A_k =-Q_k,~~~k=1,2,3.
$$
We use the energy integral $I= g_{\alpha\beta}({dX^\alpha}/{d\lambda})({dX^\beta}/{d\lambda})$
instead of the equation for the zeroth component.
Spatial components
non-canonical momenta depend only on time: $P_k = eA_k / c + Q_k$. Zeroth momentum component
also depending only on the variable $t$, it is determined from the energy condition
$g^{\mu \nu} P_\mu P_\nu = m^2c^2$.
The equations of motion will then take the form:
\begin{equation}
\frac{dX^\mu}{d\lambda}=-\frac{\sqrt{I}}{mc}g^{\alpha \mu}(X^0)P_\alpha.
\label{EuLagspecial}
\end{equation}
Eliminating
the variable $\lambda$ from here, by dividing
the expressions for the three integrals of motion by the equation for $k = 0$, we obtain
$$
\frac{dX^k}{dX^0}=\frac{g^{\mu k} (X^0) P_\mu (X^0)}{g^{\nu 0} (X^0)
 P_\nu (X^0)}=\frac{g^{\mu k} (X^0) (eA_\mu(X^0)/c+Q_\mu)}{g^{0 \nu} (X^0) (eA_\nu(X^0)/c+Q_\nu)}.
$$

{\it Example 3.}\,The generalized De Sitter Universe (a special case of Example 2):
$$
ds^2=c^2 dt^2 -\exp(2Ht) (dx^2+dy^2+dz^2)=c^2 dt^2 -\exp(2Ht)\big(dr^2+r^2 d\theta^2+r^2 \sin^2\theta d\varphi^2\big).
$$
We get $g^{\alpha \beta} = {\rm diag} (1, -e^{-2Ht}, -e^ {-2Ht}, -e^{-2Ht})$, so we have a simplification
last formula for Example 2:
$$
\frac{dX^k}{dX^0}=-\frac{P_k \exp(-2Ht)}{P_0}=-\frac{Q_k \exp(-2Ht)}{Q_0},~~~k=1,2,3.
$$
Moreover, the canonical momenta $Q_k$
are (conserved) integrals, and $Q_0$ is determined from the energy condition of Example 2:
$$
Q^2_0-\exp(-2Ht){\bf Q}^2=m^2c^2,~~~{\bf Q}^2 =Q_1^2+Q_2^2+Q_3^2,
$$
i.e. $Q^2_0 = m^2 c^2 + \exp(-2Ht){\bf Q}^2$,
$$
\frac{dX^k}{dt} \equiv V^k = \frac{c Q_k  \exp(-2Ht)}{\sqrt{m^2c^2 +{\bf Q}^2\exp(-2Ht)}}.
$$
These equations can be easily integrated,
so we get:
$$
X^k (t)= -\frac{cQ_k   \sqrt{m^2c^2 +{\bf Q}^2\exp(-2Ht) }}{H{\bf Q}^2}   +  C^k_X, $$
where arbitrary constants $C^k_X$ are determined from the initial conditions:
$$
C^k_X=\frac{cQ_k}{H {\bf Q}^2}\sqrt{m^2c^2 + {\bf Q}^2}+X^k_0.
$$
Moreover, the values of the integrals $Q_k$ are associated with Cauchy data
$V^k_0$:
$V^k_0 = Q_k/\sqrt{m^2c^2 + {\bf Q}^2}.$
We integrate the equations of motion on the time interval $[0, t]$ for $k = 1,2,3$, we get:
$$
X^k (t) = X^k(0) + \frac{cQ_k}{H {\bf Q}^2} \big(
\sqrt{m^2c^2 +  {\bf Q}^2}-\sqrt{m^2c^2 +\exp(-2Ht){\bf Q}^2}
\big).
$$
For massless particle should put $m = 0$:
$$
X^k(t)=X^k(0)+\frac{cQ_k}{H \sqrt{{\bf Q}^2}}(1-\exp(-Ht)), $$
$$
R(t)\equiv \sqrt{\sum_{k=1,2,3}(X^k(t) - X^k(0))^2}=
\frac{c}{H}\big(1 - \exp(-Ht)\big).
$$
The last formula coincides with the formula from \cite{12}, and we have obtained its generalization and derivation.

{\it Example 4.}\,In the   papers \cite{Gurzadyan1}--\cite{Gurzadyan3} it was established that the most
the general form of the ``power'' function corresponding to the case when spherical $3$--volume containing matter,
can be considered by an external observer as a point mass (from the principles of symmetry located in the center of a given volume),
is the following: $f (r) = A r^{-2} + Br$ ($ B \equiv \Lambda / \sigma$, $\sigma = {\rm const}$).
From this, based on the use of the ``weak field'' approximation, a conclusion was drawn
about the need to correct the shape of the coefficients of the metric of the point mass:
$$
g_{00} = (1-2A r^{-1} - B r^2/3)c^2, ~~~g_{11} = (1-2A r^{-1} - B r^2/3)^{-1}.
$$
Accordingly, the transition to the Friedmann--Lemaitre--Robertson--Walker (FLRW) metric allows using
of these considerations understand the structure and evaluate the impact
on the cosmological dynamics of dark matter and energy.
Using the post-Newtonian approximation based on the previously considered Fock metric allows
us to verify these conclusions. To do this, consider
the action for gravity in the approximation of weak relativism with the $\Lambda$--term has the following form
(in the Lagrangian representation):
$$
S^L = \sum_{{a}, {\bf q}} \int \frac{m_{a}}{2}{\dot{\bf x}}^2_{a} ({\bf q},t) \:dct\: -\:
\sum_{{a}, {\bf q}} \int m_{a} \Phi \big({\bf x}_{a} ({\bf q},t)\big)dct \: + $$
$$
+ \: \frac{2{\mathcal K}}{c^4}\int \int (\nabla \Phi)^2 d^3x dct \: +{\mathcal K} \int \int \Lambda d^3x dct
- \: \frac{2 {\mathcal K}\Lambda}{c^2} \int \int \Phi d^3x dct,
$$
\noindent
where ${\mathcal K} = -{c^3}/({16\pi \gamma}$).
We vary by particles, and obtain the equation of motion in the post--Newtonian approximation,
corresponding to the above action:
$$
m_{a} \ddot{\bf x}_{a}=- m_{a} \nabla \Phi({\bf x}_{a})
$$\noindent
(it turns out to coincide in form with the equation of classical dynamics).
We rewrite the action of $S$ in the Eulerian representation, introducing the classical distribution function
(on the $7$--dimensional expanded phase space):
$$
S^E = \sum_{a}
\frac{1}{2m_{a}}\int {\bf p}^2 f_{a} ({\bf x},{\bf p},t)d^3x d^3pdt\: -\:
\sum_{{a}} \int \Phi({\bf x},t) f_{a}({\bf x}, {\bf p},t) d^3p d^3x dt \:+\:
$$
$$
\: + \:\frac{2{\mathcal K}}{c^4}\int \int (\nabla \Phi)^2 d^3x dt
+{\mathcal K} \int \int \Lambda d^3x dct
\: - \: \frac{2 {\mathcal K}\Lambda}{c^2} \int \int \Phi  d^3x dt.
$$
The inverse transformation to the Lagrangian representation can be done by substituting
$f_{a}  ({\bf x}, {\bf p},t) =
\sum_{{\bf q}}  \delta \big({\bf x}- {\bf x}_{a} ({\bf q},t) \big)  \delta \big({\bf p}- {\bf p}_{a} ({\bf q},t) \big)$.
We vary $S^E$ by $\Phi$, and obtain the Poisson equation with the $\Lambda$--term:
$$
\Delta \Phi = 4\pi \gamma \sum_{a} m_{a}
\int  f_{a} ({\bf x},{\bf p},t)\:d^3p - \frac{1}{2}c^2 \Lambda.
$$
What gives the second term in the right--hand side? The presence of an ``effective'' external field: solving the equation
$\Delta \Phi = - \frac{1}{2} c^2 \Lambda$
can be chosen in the simplest form as $\Phi = -\frac{1}{12} c^2 \Lambda (x^2 + y^2 + z^2)$, which
leads to the ``repulsion'' of particles. What gives us this in a Milne--McCree--type solution?
From the Poisson equation we obtain
$$
\Phi= 4\pi \gamma \sum_{a} m_{a}
\int \frac{f_{a} ({\bf x}',{\bf p},t)}{|{\bf x}-{\bf x}'|}d^3p d^3x' - \frac{c^2 \Lambda}{12}
(x^2 +y^2 +z^2).
$$
We took advantage of the fact that the solution of an inhomogeneous linear equation is the sum
particular solutions and general solutions of the homogeneous equation, i.e., harmonic function.
Our choice of a particular solution is uniquely dictated by the requirement of isotropy (invariance
with respect to rotations) solutions of Friedmann and Milne--McCree.
The equation of the Milne model is replaced by
$$
\frac{\partial^2 {\mathcal R}}{\partial t^2} = -\gamma \frac{M(r)}{{\mathcal R}^2} + \frac{c \Lambda}{6} {\mathcal R}.
$$
We integrate the last equation, and obtain:
$$
\frac{1}{2}(\dot{\mathcal R}^2)-\gamma \frac{M(r)}{{\mathcal R}}-\frac{c^2 {\mathcal R}^2 \Lambda}{12}=E.
$$

Our task is to analyze a completely classical Lagrangian and propose a model that rationally explains
action of the formal $\Lambda$--term in Einstein's equations.

We give the corresponding ``Vlasov--Poisson equation with the $\Lambda$--term'' (for  ${a}$--th type particles):
$$
\frac{\partial f_{a}}{\partial t} + \bigg(  \frac{{\bf p}}{m_{a}}, \frac{\partial f_{a}}{\partial {\bf x}}      \bigg)
- \bigg(      \nabla \Phi,     \frac{\partial f_{a}}{\partial {\bf p}}  \bigg) =0, ~~$$
$$
\Delta \Phi = 4\pi \gamma \sum_{a}  m_{a}
\int f_{a} ({\bf x},{\bf p},t) d^3p -\frac{1}{2}c^2 \Lambda.
$$
The last equation for the potential $\Phi$ is explicitly related to the above function $f(r)$, since the last
in fact, in this case it is a derivative of the potential $\Phi$; in other worlds, the general form of the force function
for a gravitating ball (viewed in the far observation zone, which allows it to be identified with a point particle)
is a consequence of the variational principle for the Lagrangian of matter with a field.

So, we now see not only in the Lagrangians, but also in the equations of dynamics,
where to look for analogues of the $\Lambda$--term.

From the above expression for $S^E$, the mathematical analogy of $\Lambda'$ follows
and the ``cosmological parameter'' $\Lambda$ in post--Newtonian
approximation (it contains a dependence on coordinates and time), and the integral in $S^E$ c $\Lambda'$ is finite:
$$
\Lambda' ({\bf x},t) = \big(   {\mathcal K}  - 2 {\mathcal K} \Phi ({\bf x},t) c^{-2}  \big)^{-1} \bigg(
\sum_{a}
\frac{1}{2m_{a}}\int {\bf p}^2 f_{a} ({\bf x},{\bf p},t) d^3p  -\sum_a \Phi({\bf x},t)f_a ({\bf x},{\bf p},t)d^3p
\bigg).
$$

\section{Derivation of the Vlasov--Maxwell--Einstein equation in the $(3 + 3 + 1)$--dimensional $({\bf X},{\bf U},t)$--representations}
\label{sec:2}

Transition to velocity variables in 3-dimensional space for an equation of the Vlasov--Einstein type
usually it’s not considered in detail, since by default, by the way, it’s implicitly assumed a priori
that the form of the Vlasov equation changes insignificantly during this transition. But is that so?

General relativistic action
for a system of many particles with
with different masses $m_{a}$ and charges $e_{a}$ ($a = \overline{1,N}$):
in the presence of a gravitational and electromagnetic field can be written as follows:
$$
S= S_p  + S_{pf} + S_{ff} + S_{EH},
$$
$$
S_p = - \sum_{{a}} m_{{a}} c \int
\sqrt{g_{\alpha \beta} \frac{dX^{{\alpha}}_{a}}{d\lambda}
\frac{d{X}_{{a}}^{\beta}}{d\lambda}} \:d\lambda,
~~\:\:
S_{pf} = - \sum_{{a}}\frac{e_{{a}}}{c} \int
A_{\alpha}({\bf X}_{a})\frac{dX_{{a}}^{\alpha}}{d\lambda}\:d\lambda,
\eqno{(9)}
$$
$$
S_{ff} = -\:\frac{1}{16 \pi c} \int F_{\alpha \beta} F^{\alpha \beta} |g|^{1/2}\:d^4X,~~\:\:
S_{EH}=  {K} \int |g|^{1/2} ({R}+\Lambda)  \:d^4X,~~~~
$$
$$
A_\mu ({\bf X}) \equiv \{ \varphi ({{\bf X}}); {\bf A}({\bf X}) \}, ~~~
{K} = \frac{-c^3}{16 \pi \gamma},~~~{\bf X} = \{ X^\mu \}_{\mu = 0,...,3},
$$
\noindent
where: $g_{\mu \nu} ({\bf X})$ is the fundamental tensor of $4$--dimensional space--time,
$A_\mu ({\bf X})$ is  $4$--potential of the electromagnetic field,
${\Lambda}$ is a cosmological constant;
variable $\lambda \in {R}^+$
is proportional to
proper time of particle, i.e.  affine parameter of $a$--th particle:
 $$ds_a = \sqrt{I_a} d\lambda, ~~~I_a \equiv \big (g_{\mu \nu}
(dX^\mu / d \lambda) (dX^\nu / d \lambda) \big)_a$$
(${I_a}$ is a conserved integral of $a$--th particle motion).
We note specifically that the components of the metric tensor do not depend on the parameter
$\lambda$  explicitly, but only through the internal
functional dependency $4$--coordinates ${\bf X} (\lambda)$.

We obtain the equations of motion of charged massive particles in given fields by varying $S_p + S_{pf}$
(for an individual particle, the index $a = a_0$
  do not write out):
$$
-mc \frac{d^2}{d\lambda^2} \bigg(
\frac{g_{\alpha \mu} {(dX^\mu/d\lambda)}}{\sqrt{I}}
\bigg)-\frac{e}{c} \frac{dA_\alpha}{d\lambda} = -\frac{mc}{2\sqrt{I}}\frac{\partial g_{\mu \nu}}{\partial X^\alpha} \frac{dX^\mu}{d\lambda}
\frac{dX^\nu}{d\lambda}- \frac{e}{c}\frac{\partial A_\mu}{\partial X^\alpha}\frac{dX^\mu}{d\lambda}.
$$
Considering that the value $I$ is an integral of motion, we obtain the following equation:
$$
\frac{d^2 X^\mu}{d\lambda^2} + \Gamma^\mu_{\alpha \beta}\frac{dX^\alpha}{d\lambda}\frac{dX^\beta}{d\lambda}=
\frac{e\sqrt{I}}{mc^2}F_\alpha^\mu  \frac{dX^\alpha}{d\lambda},~~~\alpha,...,\mu=0,...,3.
$$
This equation is similar to equation (90.7) from the Landau and Lifshitz textbook \cite{Landau2} and
differs in the use of the parameter $\lambda$, which leads to the emergence of a root from the integral $I$.
Equations in a similar form can be found in the works of most authors who tried to derive the general relativistic Vlasov equation.
The difference, however, is that we use an arbitrary parameter $\lambda$ instead of
 affine parameter $s$ as well
the presence of the integral $I$ in the equation (when  we transfer from $d \lambda$ to $ds$ this integral is not involved).

We consider the following problem: rewrite equation (2), excluding the  parameter $\lambda$, and transferring instead of it
to the coordinate $X^0 \equiv ct $, i.e.  to ``observer time'' (which actually means refusing to use
associated with the given $a_0$--th particle of the  reference frame and the proper
time of this particle $\tau_{a_0} = ds_{a_0} / c$).
For this, initially the equations of dynamics in velocity variables:
$$
\frac{dX^\mu}{d\lambda}=V^\mu,~~~\frac{dV^\mu}{d\lambda} =- \Gamma^\mu_{\alpha \beta} V^\alpha V^\beta +
\frac{e\sqrt{I}}{mc^2}F^\mu_\alpha V^\alpha.
\eqno{(10)}
$$
We note here the appearance of the integral $\sqrt{I}$ in the second term of the right--hand side of the
second equation. It will not be when using the natural parameter
$s$ instead of $\lambda$. However, when $s$ is introduced into consideration,
the second--order homogeneity with respect to the velocities of the right--hand side disappears
the part that is needed for further conversion.
Namely, the following very general assertion holds about lowering the order by two degrees.

{\it Lemma 2} (on lowering the order of an ODE system). Let a system of $2N$ ordinary differential equations be given:
$$
\frac{dX^a}{d\lambda}=f^a  ({\bf X},{\bf V}),~~~\frac{dV^a}{d\lambda} =F^a  ({\bf X},{\bf V}),~~~a=\overline{0,N-1}.
$$
Let the functions $f^a ({\bf X},{\bf V})$ be the first degree of homogeneity with respect to
the variable ${\bf V}$, and the functions $F^a ({\bf X},{\bf V}) $ be the second degree of homogeneity:
$$
f^a  ({\bf X},k{\bf V})= k f^a  ({\bf X},{\bf V} ),~~~F^a  ({\bf X},k{\bf V})= k^2 f^a ({\bf X},{\bf V}).
$$
Then the system of $2N-2$ equations is valid:
$$
\frac{dX^a}{dX^0}=\frac{f^a ({\bf X},{\bf U})}{f^0  ({\bf X},{\bf U})},~~~
\frac{dU^a}{dX^0}=\frac{F^a  ({\bf X},{\bf U})}{f^0  ({\bf X},{\bf U})}-U^a \frac{f^a  ({\bf X},{\bf U})}{f^0  ({\bf X},{\bf U})},
$$
$$
U^a \equiv \frac{V^a}{V^0},~~U^0 \equiv 1,~~a=\overline{0,N-1}.
$$
{\it Proof} is carried out by direct substitution.

Using this Lemma, we rewrite the system (10) in the form
$$
\frac{dX^i}{dt}=U^i,~~~\frac{dU^i}{dt}=G^i({\bf X},{\bf U}),~~~i=1,2,3,
$$
$$
G^i({\bf X},{\bf U}) = -\big(    \Gamma^i_{\mu \nu}-\frac{U^i}{c}   \Gamma^0_{\mu\nu}     \big)U^\mu U^\nu
+ \frac{e\sqrt{I}}{mc^2}\big(   F^i_\mu -\frac{U^i}{c}F^0_\mu         \big)U^\mu,~~~
I \equiv g_{\mu\nu}\frac{dX^\mu}{dt}\frac{dX^\nu}{dt}.
$$
Similar equations for geodesics, only in the absence of electromagnetic fields, are given in \cite{Landau2}.

Let   us  write  the Liouville equation for $(3 + 3 + 1)$--dimensional
distribution functions $f ({\bf x}, {\bf u}, t)$ corresponding to the system (10) (hereinafter
${\bf x}\in {R}^3$, ${\bf u}\in {R}^3$, ${t}\in {R}^1$, причем ${\bf X}=\{ ct, {\bf x} \}$,
${\bf U}=\{ 1, {\bf u} \}$):
$$
\frac{\partial f({\bf x},{\bf u},t)}{\partial t}+u^i  \frac{\partial f}{\partial x^i} + \frac{\partial (fG^i)}{\partial u^i}=0.
\eqno{(11)}
$$
Thus, we obtained the first part (kinetic) of the Vlasov--Maxwell--Einstein system of equations.
To obtain the equations for the fields $g_{\mu \nu} $ and $F_\nu^\mu$
and associate these characteristics of the fields with
distribution function $f({\bf x},{\bf u},t) $,
rewrite the total action
replacing the ``arbitrary parameter'' $\lambda$ with the time $t$,
and including in $S_p$ and $S_{pf}$
partial single-particle distribution function $ f_a ({\bf x}, {\bf u}, t) $:
including in $ S_p $ and $ S_ {pf} $
partial one--particle distribution function $f_a ({\bf x}, {\bf u}, t)$:
$$
S = -\sum_a m_a c^2
\int
\sqrt{g_{\alpha \beta} U^\alpha  U^\beta}f_a({\bf x},{\bf u},t)  \:d^3x dt d^3u
 -
\sum_{{a}}\frac{e_{{a}}}{c} \int
A_{\alpha}({\bf x}_{a})  U^\alpha f_a({\bf x},{\bf u},t) {d^3  x}dt d^3 u
 -$$  $$-\frac{1}{16 \pi c} \int F_{\alpha \beta} F^{\alpha \beta}
|g|^{1/2}\:d^3 x dct +
  {\mathcal K} \int |g|^{1/2} ({R}+\Lambda)  \:d^3 x dct.  
$$

We vary the last expression for $S$ by the potentials of the electromagnetic field, and obtain the Maxwell equations:
$$
\frac{1}{8\pi} \frac{\partial
(\sqrt{-g} F^{\alpha \beta})}{\partial  X^\beta} = \sum_a \int f_a ({\bf x},{\bf u},t)U^\alpha d{\bf U}.
\eqno{(12)}
$$

We vary the action $S$ by the metric $g_{\mu \nu} $, and obtain the Einstein equations for the gravitational field:
$$
R^{\alpha \beta}- \frac{1}{2} (R+\Lambda) g^{\alpha \beta}
=-\sum_a \frac{m_a c}{{\mathcal K}\sqrt{-g}} \int \frac{U^\alpha U^\beta}{\sqrt{g_{\mu \nu} U^\mu U^\nu}}
f_a({\bf x},{\bf u},t)d{\bf u}- \frac{1}{32\pi c {\mathcal K}}
F_{\mu \nu}F^{\mu \nu} g^{\alpha \beta}.
\eqno{(13)}
$$
The system of equations (11)--(13) is the complete Vlasov--Maxwell--Einstein system.

\section{The meaning of the cosmological $\Lambda$--term}
\label{sec:3}

We see that the physical impact of the $\Lambda$--term (or, in other words, the
contribution to the $\Lambda$--term) can produce
the first three terms of the action  $S$. This obviously implies the conclusion
that the first three terms make the same contribution to the energy--momentum tensor and to the
equations of motion as the ``formal'' $\Lambda$--term:
$$
\Lambda_S  ({\bf X},t) = -\sum_a \frac{m_a c}{K \sqrt{-g}}\int \sqrt{g_{\mu \alpha}U^\mu U^\alpha}  f_a({\bf X},{\bf U},t)d^3 U -$$ $$-
\sum_a \frac{e_a}{c^2 \sqrt{-g} K} \int A_\alpha U^\alpha  f_a({\bf X},{\bf U},t) d^3 U -\frac{1}{16\pi c K}
F_{\mu\nu} F^{\mu\nu}.
$$
The notation $\Lambda_S$ emphasizes that this expression is an analogue of $\Lambda$--term, due to the form of action of $S$.
The second and third terms on the right side are associated with electromagnetism and are not sign--definite.
However, the first term is strictly positive,
since $K<0$.
For the case of a weakly relativistic metric, we have
$$\sqrt{g_{\alpha \beta} U^\alpha U^\beta} = c \sqrt{1+ 2\Phi/c^2 -U^2/c^2}$$
(here $\Phi$ is the Newtonian potential).
Therefore, the main contribution to the ``total'' $\Lambda$ is the term of the first term in $\Lambda_S$:
$$
\Lambda_{S1} ({\bf X}) = -\sum_a \frac{m_a c}{K \sqrt{-g}}\int f_a({\bf X},{\bf U},t) \sqrt{g_{\alpha \beta} U^\alpha U^\beta}d^3 U =
$$
$$
= \sum_a \frac{16 \pi \gamma m_r}{c\sqrt{-g}}\int f_a ({\bf X},{\bf U},t) \sqrt{1+2\Phi/c^2-U^2/c^2}d^3U.
$$
In fact, we can say that we have a (sign--defined) contribution to the dark energy from the distributed
(with the partial distribution function $f_a$) the rest mass $m_a c^2$.

\section{Conclusion}
\label{sec:4}

In this article, the authors followed the papers \cite{pershin1}--\cite{pershin2}, \cite{dop1}--\cite{dop3}.
The derivation of the system of Vlasov--Maxwell--Einstein equations based on the Lagrangian formalism is considered;
as the initial stage, the composite action of a system of massive charged particles, electromagnetic and gravitational fields was introduced.
The standard output option was tested using $(4 + 4)D$--space of coordinates and
velocities and its difficulties and disadvantages are explained. A new option was proposed using $(3 + 3)D$
coordinates and obtained a new form of equations. In this case, new options for action
through the distribution function, the expression of the energy--momentum tensor (on the right--hand side of the Einstein equation)
and the equations of charge motion in the general theory of relativity in the Weinberg--Fock form.
To do this, it was necessary to synchronize the intrinsic times of various particles.
We can made this by two ways: through the proper time of one particle and through an arbitrary parameter.
We derived the equations and obtained the expression for mass in stationary
gravitational and electromagnetic fields. We got solutions that depend only on time.
It is interesting to compare the obtained form of the Vlasov--Maxwell--Einstein equations with other versions and
classify them. As a rule, they are written out only for the Vlasov--Einstein equations (without Maxwell) and
with Christoffel symbols, and therefore not for impulses, but for speeds. They can also be
deduced by this scheme. In general, it is surprising that equations of the Vlasov type are not derived, but are written immediately.
This fact leads to inaccuracies. When it comes to the Vlasov--Einstein equations, the conclusion seems necessary for
both parts of the Vlasov equation: Liouville equations and equations for fields. When deriving the Liouville equation, this
led to time synchronization. In the equations for fields without derivation, the energy--momentum tensor must be taken
arbitrarily. We obtained expressions when we pass to the distribution functions in the composite action
for a system of particles in the gravitational and electromagnetic fields that
formally have the same effect as the cosmological $\Lambda$--term. We got the intimacy of the dark
energy with the rest energy of Einstein.
It seems promising to research for this
equations are all the classic permutations that are known in the Vlasov equation: energy
and hydrodynamic as well as stationary. It seems
an urgent and interesting task to classify all decisions that depend on
time (spatially homogeneous solutions). This leads to cosmological solutions that are now being actively studied.
Here the methods of the Hamilton--Jacobi equation would be useful. A very important
problem is to obtain for the equations of the Vlasov type, a statement of the type ``time averages coincide with Boltzmann extremals''.


%
%
%

%
%

\end{document}